# Future Exoplanet Research: Science Questions and How to Address Them

Jean Schneider[1]

(1)LUTH, Observatoire de Paris, PSL Research University, CNRS, Université Paris Diderot, 92190 Meudon, France

**Jean Schneider**
Email:  Jean.Schneider@obspm.fr

## Abstract

Started approximately in the late 1980s, exoplanetology has up to now unveiled the main gross bulk characteristics of planets and planetary systems. In the future it will benefit from more and more large telescopes and advanced space missions. These instruments will dramatically improve their performance in terms of photometric precision, detection speed, multipixel imaging, high-resolution spectroscopy, allowing to go much deeper in the knowledge of planets.

Here we outline some science questions which should go beyond these standard improvements and how to address them.

Our prejudice is that one is never too speculative: experience shows that the speculative predictions initially not accepted by the community have been confirmed several years later (like spectrophotometry of transits or circumbinary planets).

## Introduction

This handbook deals with all aspects of exoplanets, from pure theory to observations and instrumentation. In this section we will present speculations for the future in all these domains. Whatever the timescale of the future is, past experience shows that exoplanetology provides unexpected surprises (e.g., the unexpected discovery of 51 Peb b); predicting the future is therefore an impossible task. We will rather compile what is desirable and what means we should develop to make our dreams a reality.

With the increasing number of planets, instruments, and their observational performances, there is no doubt that we will have a more precise pictures of exoplanets. We will focus on new concepts rather than on the standard incremental aspects.

When will the "future" start? In the present section, by "future" we mean all yet not addressed or speculative science questions, audacious instrumental ideas, and yet not approved instrumental projects. What is the time horizon for the "Future of Exoplanets"? Contrary to the "No Future" pessimistic slogan, one may already design a credible road map for the whole twenty-first century and somewhat beyond. And we will take inspiration from the famous Feynman talk in 1959 "There is plenty of room at the bottom" (Feynman *1992*, *1993*) to put no restriction on the speculations, as long as they are compatible with the laws of physics. As is well known, Feynman's speculations about future manipulations of individual atoms have become a reality 50 years later with the exploding nanotechnologies.

Another example of realistic highly anticipatory vision is given by this comment by Kepler to Galileo's *Sidereus Nucius*: "There will certainly be no lack of human pioneers when we have mastered the art of flight… Let us create vessels and sails adjusted to the heavenly ether, and there will be plenty of people unafraid of the empty wastes. In the meantime, we shall prepare for the brave sky-travelers maps of the celestial bodies. I shall do it for the moon, you Galileo, for Jupiter" (Kepler *1610*).

In another words, we estimate that dreams may be productive and we will emphasize them. One can therefore consider that the anticipations presented in the present chapter do not represent a risk but an encouragement.

Of course, it is impossible to be exhaustive and, with new discoveries and instruments, new ideas will appear in the future.

Note finally that there may be different opinions, either within the present handbook or elsewhere, on some subjects, particularly regarding the touchy questions on the definition of Life and the validity of biosignatures.

In this chapter we will develop three aspects: new science questions, new objects, new instrumentation. We will also go from safe anticipation to more risky speculations, bearing in mind that "Most major discoveries in astronomy are unplanned" (Norris *2016*).

# New Science Questions

We present here some science questions and speculations from the literature and our own speculations. We arrange them according to four categories.

*(a)*
Planet properties (interior, surface, atmosphere, environment [moons, disk, tails])

*(b)*
Planet interactions with stars, cosmic rays, collisions

*(c)*
Global statistics: number of planets per star, correlation with position in the Galaxy, with stellar type and metallicity

*(d)*
Biology

Prior to any detailed aspect, a general question arises: "Which objects deserve to be qualified as planets?" The problem comes from the high mass edge of the mass histogram with the existence of 20–60 $M_{Jup}$ objects which are supposed to be formed by a mechanism different from the planet formation, namely brown dwarfs. At the lower mass limit, how to separate planets from planetesimals? There is no satisfying current solution to this problem. According to the discussion presented in the chapter *"Definition of Exoplanets and Brown Dwarfs,"* here we call "planet" any individually identified object below 60 Jupiter mass.

## Planet Properties (Interior, Surface, Atmosphere, Associated Bodies like Moons, Disks, Tails)

Let us go from the planet interior to the most external regions.

- What is the planet internal structure? Does a given giant planet have a central solid core? What are its size and mass?
- What are the planet magnetic properties?
- What are the surface features for rocky planets? Are there oceans, continents, icy surface feature (glaciers), mountains? What is the planet tectonic activity (Stamenkovic and Seager *2016*)?
- Do exoplanets have volcanoes? What are the chemical species they eject?
- What are the chemical and physical characteristics of the planet atmosphere? What is the excitation state of molecules? Are there complex molecules like C60 (Brieva et al. *2016*)? What is the thermal structure (horizontal and vertical) of the atmosphere? What is its dynamics (e.g., weather patterns, winds (Snellen et al. *2014*), vortices, storms)?
- Tides are induced by the parent star, and for exo-moons, by the planet. They deform the planet and its atmosphere and heat the planet. Observations of this deformation and heating will constrain the mechanical and thermal properties of planets.
- What are the planetary surroundings? Is there circumplanetary dust (Szulagyi et al. *2016*; Kennedy and Wyatt 2011). or are there moons? Does it have a cometary tail? (Mura et al. *2011*). For the presence of tails one can for instance rather safely speculate that for Europa-like icy small planets on an eccentric orbit there will be a water vapor tail when the planet is located in the evaporation zone of the star.
- An additional question is the planet rotation period. As already pointed out by Ford et al. (*2001*), it may (To be cautious, in the rest of the text we generally use the word "may," although the speculations presented here can often be presented as firm predictions) be inferred from the period of the modulation of the planet flux due to surface inhomogeneities.

## Planet Interactions with their Environment (Other Planets, Stars, Cosmic Rays, Collisions, Interstellar Medium)

Here we go from the most immediate to the farthest environment.

- What is the effect of the stellar wind on the planet, beyond the already well-investigated (Kotera et al. *2016*) atmosphere evaporation? For instance, similar to the delayed reaction of Solar System giant planets to solar storms (Prangé et al. *2004*; Vitasse et al. *2017*), the delayed reaction of exoplanets to stellar storms will give the speed of the stellar wind.
- Conversely, very close-in planets may induce spots on the parent star surface (Shkolnik et al. *2003*). It will be interesting to locate the position of these spots and follow their evolution.
- Past nearby supernovae explosions are likely to have had an effect on the Earth (Thomas et al. *2016*). For planets located near future exploding supernovae, may it be possible to catch the effect of these explosions on the planet atmosphere?

## Global Statistics

Global statistical questions deal with individual planetary systems and with distribution of planets in galaxies.

- A reliable statistics of abundances of the different types of planets versus stellar type is still missing, with the presence of planets around massive stars being the least known.
- While in the Solar System there are 10 objects more massive than 0.1 Earth mass (or 11 if the hypothetical Planet 9 exists), it will be interesting to see if there are planetary systems with more than 11 objects more massive than 0.1 Earth mass. Given that the more numerous planets are in a given system, the more unstable it is, it will be interesting to see what maximum number of planets around a given star can exist for a given stellar type and system architecture.
- A yet poorly known aspect is the distribution of planets in the Galaxy, their correlation with galactic arms and clusters (Brucalassi et al. *2016*), distance from the galactic center.
- Are there planets in other galaxies? What is the correlation of the planet statistics with the type of galaxy (elliptic versus spiral)?
- One can also anticipate the total number of planets that will be detected in the coming decades: while the radial velocity method will not detect more than a few thousand planets, Gaia (astrometry), Euclid (microlensing), TESS and PLATO (transits) should provide a few 10,000 planetary objects. Beyond these harvests, around 2030, the increase in the number of new planets may slow down.

# Extrasolar Life

Questions about biological or "intelligent" extrasolar life rest on prejudices about the meaning of the word "life" (Schneider *2013*). They deal with three aspects. (1) What type of life may exist? (2) On which objects may it exist? (3) How to detect it (biosignatures)?

### *What Type of Life?*

The standard prejudice is that life is nothing else than complex, probably carbon-based, chemistry, developing in liquid water. It leads to the notion of a Habitable Zone (HZ) of the parent star where the planet temperature is compatible with liquid water. But phenomenologically (in the sense of the philosopher Edmund Husserl), "life" refers to our relation sufficiently rich with objects to allow us to attribute them some autonomy, on which we project our own sense of intentional acts. In this sense "life" will perhaps be attributed to systems not based on organic chemistry. A famous example is the fictive *Black Cloud* where Fred Hoyle ( *1957*) describes the complex, and even intelligent, behavior of an interstellar complex plasma cloud. More generally, the investigation of alien forms of life should benefit from the plentiful development of the study of self-organization, even in the inorganic domain (see, e.g., Evers et al. *2016*; Tritschler and Cölfen *2016*; Wong et al. *2016*; Tsytovich et al. *2007*). Another aspect is the global properties of living systems, whatever their physicochemical substrate. One can ask if they are necessarily multicellular (on Earth an example of macroscopic unicellular organisms is already provided by the macroscopic protist *Physarum polycephalum* (Saigusa et al. *2008*) also called "blobs") or if equivalents to "animals" and "birds" might take their energy from photosynthesis. For a general discussion, see the chapter "Future Exoplanet Research: Exotic Forms of Life" in the present handbook and Orgogozo ( *2016*).

### *On Which Objects May One Find Life?*

Even for standard carbon-based organic chemistry, life can, and will perhaps, be found on objects very different from a terrestrial planet orbiting a main sequence star. A few examples have already been proposed: the moons of giant planets, such as Europa or Enceladus in our solar system, planets around K stars (Cuntz and Guinan *2016*), pulsar planets (Patruno and Kama *2017*), primordial planets (Loeb *2014*), free floating planets, and cool brown dwarfs.(Yates et al. *2017*). A brown dwarf with a surface temperature of 225–250 K has already been detected (Zapatero-Osorio et al. *2016*), allowing for the presence of liquid water. The high negentropy (the ability to increase organization) source required for out of equilibrium phenomena believed to be essential for life could be provided by the hot central temperature of the brown dwarf.

*Signatures of Life*

Since the suggestion by Lovelock (1975) that "alien biospheres" may be recognized by their gaseous dejecta, there has been slow progress in the field, mainly by the proposal to detect (biogenic) oxygen (Owen *1980*) and its by-product ozone (Bracewell and MacPhie *1979*). Until recently, the only other concept put forward is the search for nonmineral colors ("vegetation"), initiated by A. Labeyrie (*1995*). Since then, the essential debate is about the robustness of $O_2$ as a biosignature (see, e.g. , Krissansen-Totton et al. *2016*; Liu et al. *2016*; Meadows *2017*; Qiu et al. *2016*). We anticipate that if oxygen or ozone is detected on a planet, this detection will trigger hard debates and proposals for abiotic mechanisms of $O_2$ production. More recently a new approach (called ELBST) has been proposed by Airapetian et al. ( *2017*). They propose that a nitrogen-rich atmosphere is one of the fundamental prerequisites for life, which might be detected through NO emission at 5.3 micron (see also chapters "Detecting Biosignatures" and "Biosignature False Positives").

Beyond biosignatures, essentially markers of an analog of photosynthesis, "technosignatures" may reveal an artificial ("industrial") activity. Such activities may be detected by direct imaging of "city lights" or heat of industrial installations, be it on the planet itself or in large structures orbiting the planet or the central star. It has also been suggested that very large photovoltaic arrays may provide artificial colors (Loeb *2017*), whereas large orbiting structures may cause specific transit shapes (Arnold *2005*).

While technosignatures are nonintentional, an old dream, called Search for Extraterrestrial Intelligence (SETI) , is to detect intentional signals. The chapter "Exoplanets and SETI" of the present handbook gives the state of the art of the search for both intentional and nonintentional signals. It is quite possible that simplistic philosophical prejudices about "alien intelligence" (Schneider *2013*) cause the present SETI strategies the wrong way to be inefficient, and it is urgent to open our minds (Cabrol *2016*). To quote Kierkegaard ( *1844*), "It is a supreme paradox of the mind to try to discover something that our mind itself cannot think."

# New Objects

Let us now speculate here about new categories of objects.

# New Types of Planets

> New types of orbits with 1:1 resonances. In addition to standard exo-moons, there may be trojan planets (Leleu et al. *2017*, with a search in Kepler mission data reported by Hippke and Angerhausen *2015*), "1:1 eccentric resonances" (Nauenberg *2002*), "exchange orbits" (Funk et al. *2017*). As 1:1 resonances, the planet pair's radial velocity and astrometric curves limited to a few years may mimic a single planet, with the perceived mass depending on the relative position of the planets. If one of the planets is transiting, its time of transit is different from the transit of a single planet. The same goes for binary planets with unequal masses (Cabrera and Schneider *2007*).
> New planetary constituents. Possible carbon and helium planets have been considered by Mashian and Loeb ( *2016*). The ongoing developments of new particle physics theories (strings, branes, etc.) open the theoretical possibility of planetary-mass strange matter objects Schneider et al. ( *2011*). Their radius and spectra should be very different from normal (barionic) planets.

Other possible, peculiar objects are planets around black holes, neutron stars, white dwarfs (Imara and Di Stefano *2017*).

It has been speculated that planetary-mass objects may result from the tidal disruption of stars passing nearby the black hole at the center of our Galaxy (Girma and Guillochon *2017*). While they will be hardly detected by direct imaging, they could be detected by transit or lensing.

Even more speculative would be planets that advanced cultures move away from aging stars and possibly across interstellar space, and which might be illuminated by artificial microstars.

# New Planetary Features and Configurations

A planet illuminated by its parent star creates a shadow in the direction opposite to the star. When the planet is embedded in a disk this shadow produces a local decrease of temperature and of luminosity of the disk illuminated by the star (Jang-Condell *2008*, *2009*). These effects may be detected in high angular resolution imaging of the disk in infrared and visible light. Similarly, if there are two planets in the system, the planet farthest from the star may fall in the shadow of the other planet, leading to its apparent transitory disappearance (Gaidos *2017*). A similar phenomenon has been invoked earlier for the detection of exo-moons (Cabrera and Schneider *2007*).

Luger et al. ( *2016*) and Burkhart and Loeb ( *2017*) have suggested the detection of exo-aurorae on Proxima b. Similar aurorae should be detected on other nearby planets.

Cometary tails. We anticipate that small icy Europa-like planets will orbit at distances where water vapor and other gases will escape from the planet and by pushed by the stellar radiation in the form of a cometary-like tail. Such a tail, although very tenuous, has already been detected for Venus (Russell et al. *1985*). A dust-tail from a transiting low-mass planet has also been reported by Rappaport et al. ( *2012*).

High-speed winds have been detected in the upper atmosphere of beta Pic b (Snellen et al. *2014*). Whatever the explanation (fast rotating planet or rapid wind), similar features should be detected on other bright planets.

When a star is erupting, it sends light flares at the velocity of light and stellar winds at a few hundred km/s out through the whole planetary system. They should trigger a delayed excess of the illumination of the planets and atmospheric events from interaction of the stellar wind. The measurement of the time delay of the latter would give a direct measurement of the stellar wind speed. Similar observations have already been done in the Solar System (Vitasse et al. *2017*).

# Other Objects

*Moons.* The interest of exo-moons has been pointed out as soon as 1997 (Williams et al.). Although their detectability has been proven since 1999 (Sartoretti and Schneider *1999*), no secure exo-moon besides a candidate around the planet Kepler 1625 b (Teachey et al. *2017*) has been detected yet, which is an intriguing point since there are more than 60 moons in the Solar System. They present a large variety of astrophysical and biological interests, and their detection, which will indubitably occur in the near future, will help to asses some of their properties such as their mass, oblateness, habitability, etc. (Schneider etal. *2015*).

*Small bodies.* There is no doubt that small bodies , analog to similar Solar System objects, namely comets and asteroids, should exist in other planetary systems. It is also likely that some of them exist in the interstellar medium, after their ejection from a planetary system (Engelhardt et al. *2017*).

## Planet 9, A Super-Earth Analog?

Batygin and Brown ( *2016*) have suggested the existence of a new planet in the outer Solar System (called P9), supposed to be responsible for the perturbation of eccentric orbits of small bodies. Its mass being estimated between 10 and 30 Earth mass, it would constitute an excellent proxy for an extrasolar super-Earth and it is urgent to search for it.

## Known Individual Objects

In addition to new categories of objects, some already known planets present an interest for future observations:

- Proxima Centauri b (Anglada-Escudé et al. *2016*). It is not impossible that this planet is "habitable" (Dong et al. *2017*). Its main interest is its proximity to the Earth, making it the most interesting target for future interstellar probes. In the meantime, it will be the object of direct spectro-imaging with future extremely large telescopes.
- The other potential nearest planetary system is alpha Cen A,B. A controversial, but still not officially withdrawn, planet is alpha Cen B b (Dumusque et al. *2012*; Rajpaul et al. *2016*). It is not in the habitable zone of alpha Cen B, but formation scenarios give hope that there can be a habitable planet in the alpha Cen A,B system (Guedes et al. *2008*), making it a promising target for intense radial velocity, astrometric and direct imaging searches.
- The star beta Pic experienced a dimming in 1981. It was speculated that a planet was transiting the star (Lecavelier des Etangs et al. *1995*). Later in 2008 a planet was detected by imaging (Lagrange et al. *2008*). From multiple images of the planet one can infer an orbital period of about 36 years with an uncertainty of months. Therefore, if the dimming of 1981 was real and due to that planet, it must be seen again in 2017–2018, if it exists. The next transit should reappear in 2055. With the 1981 and 2017–2018 transits one should have a precise period, giving a much better precision of the 2055 transit time. At that time, it will thus be possible to devote telescope time with extremely large telescopes to catch the transit with spectrographs, allowing to explore the surroundings (atmosphere? dust?) of the planet.
- The star HD 179949 has an activity attributed to a spot induced by its hot planetary companion (Shkolnik et al. *2003*). It will be interesting to follow this spot by direct imaging of the star surface with very high angular resolution telescopes such as the Stellar Imager (Christensen-Dalsgaard et al. *2011*).
- Kepler-413 b was the first circumbinary transiting planet (Kostov et al. *2014*) which confirmed an old prediction: due to the precession of their orbits, their transits should periodically disappear and reappear (Schneider *1994*). The transits of Kepler-413 b disappeared in 2010 and should reappear for a few years around 2020 and again in 2031 (Kostov et al. *2014*). The same phenomenon has been predicted for several more circumbinary planets Martin ( *2017*).

Fomalhaut b remains an intriguing object. If its brightness is due to the reflected light of the parent star, it would have a radius 20 times the Jupiter radius. Perhaps is it surrounded by a cloud or ring of dust (Kalas et al. *2008*).

The star KIC 8462852 shows strange irregular dimming (Lisse et al. *2015*; Boyajian et al. *2016*). Several explanations have been proposed for this unusual dimming. For instance, it has been speculated that it can be due to swarms of comets (Bodman and Quillen *2015*) or to trojan asteroids. For the latter case, they could come back in 2021 (Ballesteros et al. *2018*). Intriguing periodicities have been suggested (Kiefer et al. *2017* and Sacco et al. *2017*), but with two different noncommensurable periods (928 and 1574 days)! It will be interesting to take direct images to search for these comets or trojan asteroids with very large telescopes. A follow-up of this nonstandard object can be found at http://www.wheresthefux.com/

Of course, other very interesting objects may arise in the coming years.

# How to Go Forward

Once the science questions about planets have been identified, what kind of instrumental progress do we need to address them? Most instrumental performances are not specific to exoplanets but are useful for their investigation. We successively address the principles of the detection methods, the required technical developments, and their implementation of future projects beyond approved instruments and space missions. A special case is very high contrast imaging whose domain of application is almost entirely on planets.

# Methods

Here we anticipate how future (and futuristic) new developments of observation and detection methods will help to address the above science questions: transits, radial velocity, astrometry, monopixel and multipixel imaging, radio-exploration, high resolution spectroscopy and photometry, in situ observation.

*Transits:* The transit method will go with its pace with missions like TESS, Cheops, JWST, and PLATO. One of the foreseeable improvements will come from multiple observations of secondary transits for a given transiting planet (mainly with JWST and Metis at the E-ELT) which will reveal surface features and thermal characteristics. JWST, the upcoming very large ground telescopes, as well as the ESA ARIEL space mission (Tinetti et al. *2016*, if finally selected), all employing spectrophotometry of transiting planets, are expected to profoundly improve our understanding of the composition and properties of the atmospheres of these planets.

*Astrometry :* There is, as of 2018, no planned space astrometric mission after Gaia, but on the ground the SKA project will have an astrometric accuracy of the order of a microarcsecond (Fomalont and Reid *2004*). It should thus be able to detect super-Earths around nearby stars. One must nevertheless report two artefact problems with astrometry. Indeed, a planet can be mimicked by a binary star nearby the astrometric target (Schneider and Cabrera *2006*) or by a asymmetric circumstellar disc (Kral et al. *2016*).

***Radial Velocity*** *:* High-resolution spectrographs attached to large and very large telescopes (e.g., Espresso at the VLT and HIRES at the E-ELT) will provide more and more sensitive radial velocity measurements, even for faint stars. But they will have to fight with the spurious signals of stellar activity. Progresses will have to come from the understanding of effects such as stellar surface granulation (Dumusque et al. *2017*).

*Lensing*

Observation of a given microlensing event with two or more distant telescopes, e.g., with interplanetary probes and space-ground correlations, will provide a 3D view of the events. Indeed, for large separations between telescopes the time offset between the caustic spikes will be significantly measurable. In particular, while microlensing events detected by a single telescope give only the projection of the star-planet separation on the sky plane, multisite observations will give the true star-planet separation at the time of observation. Even more, while simultaneous observations of OGLE-2015-BLG-0479 made by two telescopes (Spitzer and ground) separated by about 1 AU (Han et al. *2016* gave an offset of 13 days between two microlensing events as seen by the two observatories) probes at a distance of 5–40 AU (Pluto distance) would give one or more offsets of a month to more than 1 year. It would therefore be possible to follow the planet orbit and derive its inclination and eccentricity. One may also anticipate the detection of extragalactic planets by microlensing with large high angular resolution telescopes or interferometers (Baltz and Godolo *1999*). It would be an excellent by-product of the WFIRST mission.

Also interferometric observations of microlensing even will help to characterize in more details the microlensing events (planet mass and its projected distance to the parent star Cassan and Ranc *2016*).

Finally, a much more speculative suggestion would consist in measuring the mass of planets by the very high angular observation of their lensing effect on the cosmic microwave background (CMB). The Einstein radius is $10^9$ cm for an Earth at 30 pc. The corresponding amplification of the CMB would be observable at 1 micron with a futuristic 100,000 km space interferometer (see the chapter "Multipixel Imaging").

Mass determination of nearby planets from lensing of background sources. For planets detected by direct imaging it is not always possible to obtain their mass from radial velocity or astrometric measurements (for instance, if stellar activity is too strong). To increase the probability of a microlensing event, one has two options: increase the number of detectors in the Solar System (Zhu et al. *2016*, *2017*) or increase the number of background sources. To increase the number of detectors in the Solar System, one could take advantage of existing or planned (or dedicated) interplanetary probes. Ubiquitous background sources could simply be fluctuations of the CMB (Schneider *2018*).

***Spectro-imaging of Spatially Unresolved (Monopixel) Planets:*** In the coming decades, it may be the most productive method to characterize planets and, beyond standard characteristics, yield a rich amount of new perspectives. The exercise of future prospects of single pixel imaging is regularly done in the exoplanet community. See, for instance, the SAG15 report (Apai et al. *2017*) or the chapter "Exoplanet Atmosphere Measurements from Transmission Spectroscopy and Other Planet Star Combined Light Observations." Here we try to go beyond traditional questions.

Planet surface properties: ocean glint (Visser and van de Buit *2015*), surface polarization (Fauchez et al. *2017*), melting/freezing of ice surfaces along eccentric orbits, spectral variation due to large volcanoes (similar to Io), delayed reaction of planet to variabilities of the parent star (e.g., V404 Cyg; Gandhi et al. *2016*), temporal evolution (planet rotation, meteorology, climate, etc.).

Planet environments: rings (Arnold and Schneider *2004*), exo-moons (Cabrera and Schneider *2007*). The direct imaging of moons will give a way to infer the planet mass from the moon-planet distance and the third Kepler law. Even if the moon is not seen, it can be inferred by astrometry from accurate position variations of the parent planet. With a sufficient spectral accuracy of the planet spectrum, one may infer the presence of a moon from radial velocity variations of the planet (see the chapter "Special Cases: Moons, Rings, Comets, Trojans" in the present handbook).

Comets: it has also already been anticipated that exo-comets will be detected by imaging of their dust tail (Jura *2005*).

Catching ejections of planets from a planetary system? The simultaneous measurement of the position and radial velocity of a planet will allow to determine if it is on an escape orbit, in other words if it may become ejected from the planetary system.

The high contrast imaging approach will be improved by combining direct imaging with high spectral resolution imaging of the planet. The radial velocity of the star and its companion being very different, it will be easier to discriminate the planet from a stellar speckle (Riaud and Schneider *2007*).

**Spatially Resolved (Multipixel) Imaging of Planets and Stellar Surfaces**. Although more futuristic, this method is the most promising (Labeyrie *1999*). It will give direct access to shapes of features revealed by monopixel imaging, like oceans and continents, "forests," planetary rings, planetary transits, moon transits and shadows on the parent planet surface, volcanoes, and induced spots on the parent star surface (Shkolnik et al. *2003*).

More specifically, in the Solar System, one can infer the height of Venus mountains from the gravimetric and traveling ionospheric disturbance (turbulent regime and waves) they produce in the high atmosphere of the planet (Bertaux et al. *2016*). Multipixel imaging of exoplanets may in the future use this approach to measure the height of exo-mountains. See the chapter "Solid Exoplanet Surfaces and Relief" in the present handbook.

Industrial activity may be detected from a planet at visible light or thermal emission with unusual spatial and temporal features. Large orbiting structures may also be imaged directly.

On Earth, gigantic microbial-induced and stromatolite architectures can be detected from space (Suosaari et al. *2016*; Andrews et al. *2016*). With high-resolution multipixel imaging, analog features could be detected on nearby exoplanets.

**Other wavelengths**

**Radio Detection**. The detection of giant exoplanet's magnetospheric emission, similar to Jupiter, has been predicted since a long time (Lecacheux *1991*). But it is more efficient for hot Jupiters (Zarka et al. *2001*). It is also promising with VLBI (e.g., Katarzynski et al. *2016*). Pole-on emission, similar to W0607 + 24 (Gizis et al. *2016*), could be detected with the Chinese 500 m FAST or SKA radiotelescope. With the Next Generation VLA (ngVLA McKinnon et al. *2016*) it should be possible to peer into the internal layers of giant planet atmospheres below upper clouds, as for Jupiter (de Pater et al. *2016*). Strong molecular lines (like OH or $HC_3N$) are detected in Solar System comets at radio wavelengths (Crovisier et al. *2016*). Perhaps similar detections will be possible for extrasolar comets at their periastron for very nearby stars with the FAST radiotelescope. Air showers produced by cosmic rays in the atmosphere of our Jupiter have been investigated by Bray and Nelles (*2016*), but have been found to be undetectable with terrestrial radiotelescopes. Only in situ missions can detect them. The same holds therefore even more for exo-Jupiters. Radio exploration of star-planet interaction will reveal its impact on habitability of planetary companions, especially those in close orbits around low-mass stars (Güdel *2017*). More details are given in the chapter "Future Exoplanet Research: Radio Detection and Characterization".

*XUV*. One can safely anticipate the detection of planets by timing of X-ray pulsars or by transits. XUV will also help to investigate the star-planet interaction. Finally, perhaps X-ray imaging of planets, similar to the detection of Pluto by the Chandra mission (Lisse et al. *2016*), will be possible by the Athena mission or successors. More details are found in the "Future Exoplanet Research: XUV Detection and Characterization" chapter in the present handbook.

### High-Precision Observations

High or very high precision observations will be possible with 30 meter class telescopes or larger, thanks to the large number of collected photons. They will allow high resolution spectroscopy, high precision photometry, and fast observations.

### Very High-Resolution Spectroscopy of Planets

For instance, some industrial gases, such as chlorofluorocarbons (CFCs) have narrow spectral lines and could be detected by high-resolution spectroscopy of planets (Schneider et al. *2010*).

### High-Precision Photometry

Direct imaging with extremely large telescopes may allow to sound the interior of giant planets by seismology, similar to stellar seismology (see Gaulme et al. *2014* for solar system giant planets).

*Fast Observations*. Thanks to rapid observations it may become possible to detect rapid changes in planet characteristics. For instance, rapid multipixel imaging of stellar surfaces may allow to see transiting planets or moons in action. Presumably, in the future, in the case of technological civilizations, rapid changes may rather be due to industrial activities than to biological events.

*Laboratory Work*. The knowledge of planet interiors will benefit from the advanced studies of equations of state, with the support of experiments at very high pressure at facilities like the US National Ignition Facility (Bolis et al. *2016*). The formation of planets can be investigated experimentally by the study of grain collisions under microgravity conditions in the ISS (Brisset et al. *2017*). Further multiparticle collision experiments should be performed in the future.

**In Situ** *Observation*. While multipixel imaging should arise in the middle of the twenty-first century , it will never achieve the sufficient angular resolution to observe the morphology of organisms ("trees," "animals") even on the nearest exoplanets. Indeed, an AU-sized telescope would be required for this type of observation (Schneider et al. *2010*). Only in situ observations, requiring an interstellar travel mission, will achieve that goal (Schneider *2010*), unless we receive images sent by "extraterrestrials." Although impossible to predict, there is no scientific objection against this event to happen any time. The idea of interstellar missions has progressively shifted from science fiction to actual projects. After the initial Daedalus project, started in the 1970s (Bond and Martin *1975*) and an explicit application to exoplanets (Wolczek *1982*), the latest step forward is the Breakthrough Starshot Initiative (see for instance Heller *2017* and Hippke *2017* for an update). More details are given in the present handbook's chapter ["Direct Exoplanet Investigation Using Interstellar Space Probes"](#).

# Technology and Materials

Many technical improvements will occur in detector technology, extension of wavelength ranges, mirror surface smoothness, optics, coronagraphs. Here we speculate on some nonstandard aspects. Again it is impossible to be exhaustive.

Several planets are found in binary star systems. They present a challenge for direct imaging since presently high contrast imaging (coronagraphy or interferometric nulling) works only for a single source. There is nevertheless a recent improvement in coronagraphy which opens the hope to take images of planets in binary star systems (Sirbu et al. *2017*). This is of particular importance for the alpha Cen A,B system where there is still a hope that it contains a terrestrial planet (Guedes et al. *2008*).

Monitoring of the roughness of mirrors and various optical surface will be improved thanks to the interference of diffuse light with specularly reflected light of surfaces (Barrelet *2016*). Will we be able to push the high surface precision down to the limits of quantum fluctuations of surfaces, similarly to the philosophy of Feynman's *1993* paper "There's plenty of room at the bottom"?

Intensity interferometry will investigate planetary transits in front of the brightest stars with high angular resolution imaging (Dravins *2016*). A first successful test has been made by Guerin et al ( *2017*).

Graphene has become a quasi-magic material with many virtues. For instance, it may allow for ultra-light sails for interstellar missions (Scheffer *2015*). Other materials, converting infrared into visible, will improve detector technology (Roseman et al. *2016*).

A review of future of some of these developments is given by the chapter ["Future Exoplanet Research: High Contrast Imaging Techniques"](#) in the present handbook.

# Individual Instruments and Facilities

In the past decades, several road maps have been elaborated to recommend future instruments, like for instance the European infrared interferometer Darwin (Léger et al. *1996*) and its American brother

Terrestrial Planet Finder (Beichman et al. *1999*). Unfortunately, more than 20 years later the required technologies are still not available with low risk levels and neither in the current ESA or NASA schedules may such a mission happen before 2040. Let us nevertheless list current more or less futuristic ideas, beyond the coming soon 30 m-class optical telescopes or SKA, hoping that a few of them will become real as soon as possible.

- The Next Generation VLA, ngVLA, (McKinnon et al. *2016*) will extend by a factor 10 both the sensitivity and angular resolution of the VLA, allowing for the radio investigation of nearby exoplanets.
- Starshades are external 50 m-size coronagraphs, initially proposed by W. Cash et al. (*2003*), blocking stellar light for high-contrast imaging, located at 70,000 kilometers from a standard space telescope. In its current format it could be associated with the JWST. A starshade is also currently studied for the WFIRST mission.
- The WFIRST (Wide Field Infrared Survey Telescope) project, initially dedicated to cosmology, has recently got an extension, WFIRST/AFTA, with a coronagraph for visible wavelengths in its focal plane (Macintosh and Robinson *2016*). Its launch is foreseen for around 2025, although the final configuration is (as of October 2017) still under discussion (Zurbuchen *2017*).
- The World Space Observatory project, started in the late 1990s, is a 2.5 m-class telescope with an UV camera. It could detect the Lyman-alpha line in the atmosphere of Earth-like exoplanets (Gomez de Castro et al. *2017*.) The adjunction of a coronagraph has been recently proposed (Shashkova et al. *2017*).
- The EXCITE project (EXoplanet Infrared Climate TElescope) aims at the characterization of atmospheres in the 1–4 micron region with a 0.5 m class telescope on a long duration balloon flight (Pascale et al. *2017*).
- The Stellar Imager (Christensen-Dalsgaard et al. *2011*) is an interferometric project aimed at the imaging of bright stellar surfaces. It could take images of planetary transits caught in the action.
- It has been proposed by Heidmann and Maccone (*1994*) and Maccone (*2009*) to use the Sun as a gravitational lens to amplify the signal from exoplanets. The corresponding focus is at 550 AU from the Earth, requiring thus a mission, which they name FOCAL, at the border of the Solar System. One of the limitations of this idea is that a gravitational lens does not generate an enlarged image but projects the background object into a ring; it primarily would amplify its signal. It therefore does not solve the difficulties of high-contrast imaging. A recent critical analysis can be found in Landis (*2016*).
- LUVOIR is one of four US Decadal Survey Mission Concept Studies initiated in 2016. It is a 8–16 m space telescope project, equipped with a coronagraph (or an external starshade) aimed at the spectro-imaging of exo-Earths from 0.15 to 5 microns (Stark et al. *2015*).
- The Colossus/ParFAIT project is an ambitious 70 m class ground-based telescope for the spectro-imaging of exo-Earths (Kuhn et al. *2014*).
- The nonconventional concept of "densified pupil" (a device in the focal plane) has been proposed by Labeyrie in Labeyrie *1996*. It allows to have a clear 2D image with an interferometer consisting of many very dispersed sub-apertures. Equipped with a pupil densifier at the focus of the interferometer, the latter is called a "hypertelescope." With a baseline of thousands of kilometers in space it would provide multipixel images of exo-Earths. A possible precursor could be located on the Moon (Labeyrie *2017*). A ground-based demonstrator is presently being tested in the south Alps (Labeyrie *2016*).

Another example of a nonconventional telescope is the SPIDER project (Segmented Planar Imaging Detector for EO Reconnaissance, Kendrick et al. *2013*). It replaces large mirrors or conventional interferometers by a densely packed interferometer array based on photonic integrated circuits. Although initially not dedicated to exo-planets, it could some day be used for their detection.

## A Very Tentative Schedule of Anticipated Discoveries and Facilities

Let us finally risk a tentative and partial schedule of future events. Some benchmarks are given by the schedule of Space Agencies and International Organizations.

| Time | Event | Facility |
| --- | --- | --- |
| 2018–2020 | Exo-moons | TESS, Cheops, JWST |
| 2025–2030 | Transiting habitable planets | Plato |
| 2025–2030 | X-ray detection | Athena WFIRST |
|  | Direct imaging | Planetary Systems Camera at the E-ELT |
| 2030–2040 | Spectro-imaging of exo-Earths | 8 m space telescope or large interferometer |
| 2040–2100 | Spectro-imaging of exo-Earths | Colossus |
|  | Multipixel imaging | Hypertelescope |
| Beyond 2100 | In situ observation | Interstellar mission |

## Conclusion

The future of exoplanets is brilliant. The above considerations and the other chapters in this section can serve as a modest draft roadmap for the twenty-first century. But progress of the field is constrained by the competition with other scientific domains, given the limited budgets of agencies. Big projects could benefit from coordination between countries, for instance, in the framework of a World Exoplanet Programme (Schneider et al. *2009*).
An update of some of the speculations presented here will be found in http://luth7.obspm.fr/exo-speculations.html.
One can find also some speculations and advanced investigations in https://www.centauri-dreams.org/
https://www.nasa.gov/directorates/spacetech/niac/index.html
http://www.esa.int/gsp/ACT/publications/ActaFutura/index.html
http://www.esa.int/gsp/ACT/index.html .

# References

Airapetian V et al (2017) Detecting the beacons of life with exo-life beacon space telescope (ELBST). In: Planetary science vision 2050 workshop, held 27–28 Feb and 1 Mar 2017, Washington, DC. LPI contribution no. 1989, id.8214

Andrews J et al (2016) Exhumed hydrocarbon-seep authogenic carbonates from Zakynthos Island (Greece): concretions not archaeological remains. Mar Pet Geol 76:16
CrossRef

Anglada-Escudé G et al (2016) A terrestrial planet candidate in a temperate orbit around Proxima Centauri. Nature 536:437
ADS CrossRef

Apai D et al (2017) Exploring other worlds: science questions for future direct imaging missions (EXOPAG SAG15 Report). arxiv:1708.02821

Arnold and Schneider (2004) The detectability of extrasolar planet surroundings. I. Reflected-light photometry of unresolved rings. Astron Astrophys 420:1153

Arnold L (2005) Transit light-curve signatures of artificial objects. ApJ 627:534
ADS CrossRef

Austin R (2016) Phys Today 69(12):42
ADS CrossRef

Ballesteros F et al (2018) KIC 8462852: Will the Trojans return in 2021? MNRAS 473:L21

Baltz E, Godolo P (1999) Searching for extragalactic planets. https://arxiv.org/abs/astro-ph/9909510. http://adsabs.harvard.edu/abs/2016AAS...22821705B

Barrelet E (2016) Direct illumination calibration of telescopes at the quantum precision limit. Astron Astrophys 594:A38. arXiv:1610.00474
ADS CrossRef

Batygin K, Brown M (2016) Astron J 151:22
ADS CrossRef

Beichman C et al (1999) The Terrestrial Planet Finder (TPF): a NASA origins program to search for habitable planets. JPL Publication, Pasadena. 99–3

Bertaux J-L et al (2016) Influence of Venus topography on the zonal wind and UV albedo at cloud top level: the role of stationary gravity waves. J Geophys Res Pap 121:1087
ADS CrossRef

Bodman E, Quillen A (2015) KIC 8462852: transit of a large comet family. ApJ Lett 819:L34
ADS CrossRef

Bolis R et al (2016) Decaying shock studies of phase transitions in MgO-SiO$_2$ systems: implications for the super-Earths' interiors. Geophys Res Lett 43:9475
ADS CrossRef

Bond A, Martin A (1975) Project Daedalus: the origins and aims of the study. J Br Interplanet Soc 28:146
ADS

Boss A (1995) Proximity of Jupiter-like planets to low-mass stars. Science 267:360
ADS CrossRef

Boyajian T et al (2016) Planet hunters X. KIC 8462852 – where's the flux? MNRAS 457:3988
ADS CrossRef

Bracewell R, MacPhie R (1979) Searching for nonsolar planets. Icarus 38:136
ADS CrossRef

Bray J, Nelles A (2016) Minimal prospects for radio detection of extensive air showers in the atmosphere of Jupiter. ApJ 825:129
ADS CrossRef

Brieva AC et al (2016) ApJ C60 as a probe for astrophysical environments. arXiv:1605.08745

Brisset J et al (2017) NanoRocks: design and performance of an experiment studying planet formation on the international space station. arXiv:1706.08625

Brucalassi A et al (2016) Search for giant planets in M67 III: excess of hot Jupiters in dense open clusters. Astron Astrophys. ArXiv:1606.05247

Burkhart B, Loeb A (2017) The detectability of radio auroral emission from Proxima b. ApJ Lett. Submitted. https://arxiv.org/abs/1706.07038

Cabrera J, Schneider J (2007) Detecting companions to extrasolar planets using mutual events. Astron Astrophys 464:1133
ADS CrossRef

Cabrol N (2016) Alien mindscapes – a perspective on the search for extraterrestrial intelligence. Astrobiology 16:661. http://www.spaceref.com/news/viewpr.html?pid=49044
ADS CrossRef

Cash W et al (2003) The new worlds observer: a new approach to observing extrasolar planets. Bull Am Astron Soc 35:1416. American Astronomical Society Meeting 203, id.130.06
ADS

Cassan A, Ranc C (2016) Interferometric observation of microlensing events. MNRAS 458:2074
ADS CrossRef

Christensen-Dalsgaard J et al (2011) The Stellar Imager (SI) – a mission to resolve stellar surfaces, interiors, and magnetic activity. J Phys Conf Ser 271:012085
CrossRef

Cridland A et al (2016) Composition of early planetary atmospheres I: connecting disk astrochemistry to the formation of planetary atmospheres. MNRAS. arXiv:1605.09407

Crovisier J et al (2016) Comets at radio wavelengths. In: Proceedings of URSI France scientific days, "Probing matter with electromagnetic waves", 24–25 Mar 2015, Paris. To be published in C. R. Physique arXiv:1606.06020

Cuntz M, Guinan E (2016) About exobiology: the case for dwarf K stars. ApJ. arXiv:1606.09580

De Pater I et al (2016) Peering through Jupiter's clouds with radio spectral imaging. Science 352:1198

ADS CrossRef

Dong C et al (2017) Is Proxima Centauri b habitable? A study of atmospheric loss. ApJ Lett 837:L26
ADS CrossRef

Dravins D (2016) Intensity interferometry: optical imaging with kilometer baselines. In: SPIE 9907. https://arxiv.org/abs/1607.03490

Dumusque X et al (2012) An Earth-mass planet orbiting α Centauri B. Nature 491:207
ADS CrossRef

Dumusque X et al (2017) Radial-velocity fitting challenge. II. First results of the analysis of the data set. Astron Astrophys 598:A133
CrossRef

Engelhardt J, Jedicke R, Vereš P et al (2017) An observational upper limit on the interstellar number density of asteroids and comets. Astron J 153:133
ADS CrossRef

Evers CH et al (2016) Self-assembly of microcapsules via colloidal bond hybridization and anisotropy. Nature. https://doi.org/10.1038/nature17956

Fauchez TH et al (2017) The $O_2$ A-band in fluxes and polarization of starlight reflected by Earth-like exoplanets. ApJ 842:41
ADS CrossRef

Feng Y et al (2016) The impact of non-uniform thermal structure on the interpretation of exoplanet emission spectra. ApJ arXiv:1607.03230

Feynman R (1992) There's plenty of room at the bottom. J Microelectromech Syst 1(1):60–66. https://doi.org/10.1109/84.128057
CrossRef

Feynman R (1993) Infinitesimal machinery. J Microelectromech Syst 2(1):4–14. https://doi.org/10.1109/84.232589
MathSciNet CrossRef

Fomalont E, Reid M (2004) Microarcsecond astrometry using the SKA. New Astron Rev 48:1473
ADS CrossRef

Ford E, Seager S, Turner E (2001) Characterization of extrasolar terrestrial planets from diurnal photometric variability. Nature 412:885
ADS CrossRef

Funk B et al (2017) Exchange orbits – an interesting case of co-orbital motion. arxiv:1708.04205

Gaidos E (2017) Transit detection of a "Starshade" at the inner lagrange point of an exoplanet. MNRAS 469:4455
ADS CrossRef

Gandhi P et al (2016) Furiously fast and red: sub-second optical flaring in V404 Cyg during the 2015 outburst peak. MNRAS 459:554
ADS CrossRef

Gaulme P, Mosser B, Schmider F-X, Guillot T, Jackiewicz J (2014) Seismology of giant planets. 2014arXiv1411.1740G

Girma E, Guillochon J (2017) Modeling the spatial distribution of fragments formed from tidally disrupted stars. AAS meeting #229, id.154.13

Gizis J et al (2016) WISEP J060738.65+242953.4: a nearby. Pole-on L8 brown dwarf with radio emission. Astron J. arXiv:1607.00943

Gomez de Castro A et al (2017) On the feasibility of studying the exospheres of Earth-like exoplanets by Lyman-alpha monitoring. Detectability constraints for nearby M stars. Exp Astron. Submitted.
https://arxiv.org/abs/1704.07443

Güdel M (2017) Exoplanetary habitability: radiation, particles, plasmas, and magnetic fields. In: AASTCS5 radio exploration of planetary habitability. Proceedings of the conference 7–12 May 2017, Palm Springs. Published in Bulletin of the American Astronomical Society, vol 49, no 3, id.100.01

Guedes J, Rivera J, Davis E et al (2008) Formation and detectability of terrestrial planets around alpha Centauri B. ApJ 679:1582
ADS CrossRef

Guerin W et al (2017) Temporal intensity interferometry: photon bunching on three bright stars. MNRAS. Accepted. arxiv:1708.06119

Guggenberger E et al (2016) Significantly improving stellar mass and radius estimates: a new reference function for the Δν scaling relation. MNRAS. arXiv:1606.01917

Hamers A, Portegies Zwart S (2016) White dwarf pollution by planets in stellar binaries. MNRAS. arXiv:1607.01397

Han C, Udalski A, Gould A et al (2016) OGLE-2015-BLG-0479LA,B: binary gravitational microlens characterized by simultaneous ground-based and space-based observation. ApJ 828:53
ADS CrossRef

Heidmann J, Klein M (1991) Bioastronomy: the search for extraterrestrial life. Springer, Berlin
CrossRef

Heidmann J, Maccone C (1994) Astrosail and SETIsail: two extrasolar system missions to the Sun's gravitational focuses. Acta Astronaut 32:409
ADS CrossRef

Heller R (2017) Relativistic generalization of the incentive trap of interstellar travel with application to breakthrough starshot. MNRAS. https://doi.org/10.1093/mnras/stx1493

Hippke M (2017) Interstellar communication I. Maximized data rate for lightweight space-probes. Arxiv: 1706.03795

Hippke M, Angerhausen D (2015) A statistical search for a population of exo-trojans in the Kepler data set. ApJ 811:1
ADS CrossRef

Hoyle (1957) The Black Cloud (New edition: Penguin Classics 2010)

Imara N, Di Stefano R (2017) Searching for exoplanets around X-ray binaries with accreting white dwarfs, neutron stars, and black holes. ApJ. Submitted. https://arxiv.org/abs/1703.05762

Jang-Condell H (2008) Planet shadows in protoplanetary disks. I. Temperature perturbations. ApJ 679:797
ADS CrossRef

Jang-Condell H (2009) Planet shadows in protoplanetary disks. II observable signatures. ApJ 700:820
ADS CrossRef

Joachimi K et al (2016) On the detectability of CO molecules in the interstellar medium via X-ray spectroscopy. MNRAS. arXiv:1606.02285

Jura M (2005) The age-dependence of the detectability of comets orbiting solar-type stars. ApJ 620:487
ADS CrossRef

Kalas P et al (2008) Optical images of an exosolar planet 25 light-years from Earth. Science 322:1345
ADS CrossRef

Katarzynski K et al (2016) Search for exoplanets and brown dwarfs with VLBI. MNRAS 461:929
ADS CrossRef

Kendrick R et al (2013) Flat panel space based space surveillance sensor. In: Proceedings of the advanced Maui optical and space surveillance technologies conference, held in Wailea, Maui, 10–13 Sept 2013, Ed.: S. Ryan, id.E45

Kennedy G, Wyatt M (2011) Collisional evolution of irregular satellite swarms: detectable dust around Solar system and extrasolar planets. MNRAS 414:2137

Kepler J (1610) Dissertatio cum nuncio siderio p 39. Cited in the Moon and the western imagination by Scott Montegomery (University of Arizona Press 2001) p 121. Kepler's conversation with Galileo's sidereal messenger. Translated by E. Rosen. New York, 1965. Translation of Dissertatio cum Nuncio Siderio (1610)

Kiefer F et al (2017) Detection of a repeated transit signature in the light curve of the enigma star KIC 8462852: a 928-day period? Astron Astrophys. Submitted. arxiv:1709.01732

Kierkegaard S (1844) Philosophical fragments

Kite E et al (2016) Atmosphere-interior exchange on hot rocky exoplanets. ApJ. arXiv:1606.06740

Kluska J, Benisty M, Soulez F et al (2016) Astron Astrophys. Arxiv: 1605.05262

Kostov V et al (2014) Kepler-413b: a slightly misaligned, Neptune-size transiting circumbinary planet. ApJ 784:14
ADS CrossRef

Kotera K, Mottez F, Voisin G, Heyvaerts J (2016) Astron Astrophys. ArXiv:1605. 05746

Kral Q, Schneider J, Kennery G, Souami D (2016) Effects of disc asymmetries on astrometric measurements– can they mimic planets? Astron Astrophys 592:A39
ADS CrossRef

Krissansen-Totton J et al (2016) On detecting biospheres from chemical thermodynamic disequilibrium in planetary atmospheres. Astrobiology 16:39. https://astrobiology.nasa.gov/news/could-earths-light-blue-color-be-a-signature-of-life/
ADS CrossRef

Kuhn J et al (2014) Looking beyond 30m-class. Telescopes: the colossus project. SPIE Astronomical Telescopes and Instrumentation, 9145, id 91451G

Labeyrie A (1995) Private communication

Labeyrie A (1996) Resolved imaging of extra-solar planets with future 10–100km optical interferometric arrays. Astron Astrophys Suppl 118:517
ADS CrossRef

Labeyrie (1999) Snapshots of alien worlds – the future of interferometry. Science 285:1864

Labeyrie A (2016) Hypertelescopes: potential science gains, current testing and prospects in space. EAS Publ Ser 78–79:45

CrossRef

Labeyrie A (2017) Hypertelescope. In: Astronomy and science from the Moon workshop, June 22. Institut d'astrophysique de Paris

Lagrange A-M et al (2008) A probable giant planet imaged in the beta Pictoris disk. VLT/NACO deep L-band imaging. Astron Astrophys 493:L21
ADS CrossRef

Landis G (2016) Mission to the gravitational focus of the Sun: a critical analysis. arXiv:1604.06351

Lecacheux A (1991) On the feasibility of extra-solar planetary detection at very low radio frequencies. In: Heidmann J, Klein MJ (ed) Bioastronomy the search for extraterrestial life – the exploration broadens. Lecture notes in physics, vol 390. Proceedings of the third international symposium on bioastronomy held at Val Cenis, Savoie, 18–23 June 1990, p 21

Lecavelier des Etangs et al (1995) Pictoris: evidence of light variations

Léger A et al (1996) Could we search for primitive life on extrasolar planets in the near future? Icarus 123:249
ADS CrossRef

Leleu A et al (2017) Detection of co-orbital planets by combining transit and radial-velocity measurements. Astron Astrophys 599:L7
ADS CrossRef

Lisse C et al (2015) IRTF/SPEX observations of the unusual Kepler lightcurve system KIC8462852. ApJ Lett 815:L17
ADS CrossRef

Lisse C et al (2016) The puzzling detection of x-rays from Pluto by Chandra. Icarus. https://doi.org/10.1016/j.icarus.2016.07.008

Liu C et al (2016) Water splitting–biosynthetic system with $CO_2$ reduction efficiencies exceeding photosynthesis. Science 252:1210
ADS CrossRef

Loeb A (2014) The habitable epoch of the early universe. Int J Astrobiol 13:337
CrossRef

Loeb A (2017) Natural and artificial spectral edges in exoplanets. MNRAS (in press). https://arxiv.org/abs/1702.05500

Lovelock J (1975) Thermodynamics and the recognition of alien biospheres. Proceedings of the Royal Society of London. Series B, Biological Sciences 189:167

Luger R et al (2016) The pale green dot: a method to characterize Proxima Centauri b using exo-aurorae. ApJ. Accepted. https://arxiv.org/abs/1609.09075

Maccone C (2009) Deep space flight and communications: exploiting the Sun as a gravitational lens. Springer, Berlin. ISBN: 978-3-540-72942-6
CrossRef

Macintosh B, Robinson T (2016) Exoplanet detection and characterization with the WFIRST space coronagraph. In: American Geophysical Union, Fall General Assembly 2016. Abstract #P13C-04

Martin DV (2017) Circumbinary planets – II. When transits come and go. MNRAS 465:3235
ADS CrossRef

Mashian N, Loeb A (2016) CEMP stars: possible hosts to carbon planets in the early universe. MNRAS. arXiv:1603.06943

McGuire B, Brandon Carroll P, Loomis RA, Finneran IA, Jewell PR et al (2016) Discovery of the interstellar chiral molecule propylene oxide ($CH_3CHCH_2O$). Science 352:1449–1452
ADS CrossRef

McKinnon M et al (2016) A preliminary operations concept for the ngVLA. In: Proceedings of the SPIE, vol 9910, id. 99100L

Meadows V (2017) Reflections on $O_2$ as a biosignature in exoplanetary atmospheres. Astrobiology. https://doi.org/10.1089/ast.2016.1578. http://www.planetformationimager.org/

Mura A, Wurz P, Schneider J et al (2011) Comet-like tail-formation of exospheres of hot rocky exoplanets: possible implications for Corot-7b. Icarus 211:1
ADS CrossRef

Nauenberg M (2002) Stability and eccentricity of periodic orbit for two planets in a 1:1 resonance. Astron J 124:2332
ADS CrossRef

Nielsen J et al (2016) Eye lens radiocarbon reveals centuries of longevity in the Greenland shark (*Somniosus microcephalus*). Science 353:702
ADS CrossRef

Norris R (2016) Discovering the unexpected in astronomical survey data. Publ Astron Soc Aust. https://doi.org/10.1017/pasa.2016.63

Oklopcic A et al (2016) ApJ. Accepted. ArXiv:1605.07185

Orgogozo V (2016) Imagine living in a parallel world. J CNRS. https://news.cnrs.fr/opinions/imagine-living-in-a-parallel-world

Owen T (1980) The search for early forms of life in other planetary systems: future possibilities afforded by spectroscopic techniques. In: Papagiannis MD (ed) Strategies for the search for life in the universe. Reidel, Dordrecht, p 177
CrossRef

Owen J, Kollmeier J (2016) Dust traps as planetary birthsites: basics and vortex formation. MNRAS. arXiv:1607.08250

Paine M (2006) Can we detect asteroid impacts with rocky extrasolar planets? http://www.thespacereview.com/article/761/1

Pascale E et al (2017) The EXoplanet Infrared Climate TElescope (EXCITE) in European Planetary Science Congress 2017, vol 11, EPSC2017-729-1

Patruno A, Kama M (2017) Neutron star planets: atmospheric processes and habitability. Astron Astrophys. Submitted. https://arxiv.org/abs/1705.07688

Payne M, Veras D, Gaensicke B, Holman M (2017) The fate of exomoons in white dwarf planetary systems. MNRAS. Accepted. https://arxiv.org/abs/1610.01597

Pisano G, Maffei B, Ade P et al (2016) Multi-octave metamaterial reflective half-wave plate for millimetre and sub-millimetre wave applications applied optics. Submitted. https://arxiv.org/abs/1610.00582

Prangé R, Pallier L, Hansen K et al (2004) An interplanetary shock traced by planetary auroral storms from the Sun to Saturn. Nature 432:78
ADS CrossRef

Qiu Y et al (2016) Efficient solar-driven water splitting by nanocone BiVO4-perovskite tandem cells. Sci Adv 2(6):e1501764. https://doi.org/10.1126/sciadv.1501764
ADS CrossRef

Quintana E et al (2016) The frequency of giant impacts on Earth-like worlds. ApJ 821:126. http://www.manyworlds.space/index.php/2016/05/23/big-bangs/
ADS CrossRef

Rajpaul V et al (2016) Ghost in the time series: no planet for Alpha Cen B. MNRAS 456:L6
ADS CrossRef

Rappaport S et al (2012) Possible disintegrating short-period super-mercury orbiting KIC 12557548. ApJ 752:1
ADS CrossRef

Riaud P, Schneider J (2007) Improving Earth-like planets detection with an ELT: the differential radial velocity experiment. Astron Astrophys 469:355
ADS CrossRef

Roseman N et al (2016) A highly efficient directional molecular white-light emitter driven by a continuous-wave laser diode. Science 352:1301
ADS MathSciNet CrossRef MATH

Russell M, Saunders M, Luhman J (1985) Mass-loading and the formation of the Venus tail. Adv Space Res 5:1
ADS CrossRef

Sacco G et al (2017) A 1574-day periodicity of transits orbiting KIC 8462852. ApJ. Submitted. arxiv:1710.01081

Saigusa T, Tero A, Nakagaki T, Kuramoto Y (2008) Amoebae anticipate periodic events. Phys Rev Letters 100:018101
ADS CrossRef

Sartoretti P, Schneider J (1999) On the detection of satellites of extrasolar planets with the method of transits. Astron Astrophys Suppl 134:553
ADS CrossRef

Scheffer L (2015) Graphene sails with phased array optical drive – towards more practical interstellar probes. https://arxiv.org/abs/1506.09214

Schneider J (1994) On the occultations of a binary star by a circum-orbiting dark companion. Planet Space Sci 42:539
ADS CrossRef

Schneider J (2010) Reply to a comment on "The far future of exoplanet direct characterization" – the case for interstellar space probes. Astrobiology 10:857
ADS CrossRef

Schneider J et al (2011) Defining and cataloging exoplanets: the exoplanet.eu database. Astron Astrophys 532:A79

Schneider J (2013) Philosophical issues in the search for extraterrestrial life and intelligence. Int J Astrobiol 12:259
CrossRef

Schneider (2017) Measuring the mass and radius of Planet 9 PASP 129:104401

Schneider J, Cabrera J (2006) Can stellar wobble in triple systems mimic a planet? Astron Astrophys 445:1159

[ADS](#) [CrossRef](#)

Schneider J, Coudé du Foresto V, Ollivier M (2009) Search for life on exoplanets: toward an international institutional coordination. In: Exoplanets and disks: their formation and diversity. In: Proceedings of the international conference. AIP conference proceedings, vol 1158, p 369

Schneider J et al (2010) The far future of exoplanet direct characterization. Astrobiology 10:121
[ADS](#) [CrossRef](#)

Schneider J, Lainey V, Cabrera J (2015) A next step in exoplanetology: exo-moons. Int J Astrobiol 14:191
[CrossRef](#)

Shashkova I et al (2017) Stellar imaging coronagraph an additional instrument for exoplanet exploration onboard the WSO-UV 1.7 meter orbital telescope. In: European planetary science congress 2017, vol 11, EPSC2017-536-1

Shkolnik E, Walker G, Bohlender D (2003) Evidence for planet-induced chromospheric activity on HD 179949. ApJ 597:1092
[ADS](#) [CrossRef](#)

Simpson F (2016) An anthropic prediction for the prevalence of waterworlds. MNRAS 468:2803
[ADS](#) [CrossRef](#)

Sirbu D, Thomas S, Belikov R (2017) Techniques for high-contrast imaging in multi-star systems II: multi-star wavefront control. ApJ. Submitted. arxiv:1704.05441

Snellen I, Brandl B, de Kok R et al (2014) Fast spin of the young extrasolar planet β Pictoris b. Nature 509:63
[ADS](#) [CrossRef](#)

Stamenkovic V, Seager S (2016) Emerging possibilities and insuperable limitations of exogeophysics: the example of plate tectonics. ApJ 825:78
[ADS](#) [CrossRef](#)

Stark C et al (2015) Lower limits on aperture size for an exoearth detecting coronagraphic mission. ApJ 808:149

ADS CrossRef

Suosaari E et al (2016) New multi-scale perspectives on the stromatolites of Shark Bay, Western Australia. Sci Rep 6:20557
ADS CrossRef

Szulagyi J, Masset F, Lega E et al (2016) Circumplanetary disk or circumplanetary envelope? MNRAS. Arviv: 1605.04586

Teachey A, Kipping DM, Schmitt AR (2017). HEK VI: on the dearth of Galilean analogs in Kepler and the exomoon candidate Kepler-1625b I. arXiv:1707.08563

Thomas B, Engler E, Kachelriess M et al (2016) Terrestrial effects of nearby supernovae in the early pleistocene. ApJ Lett 826:L3
ADS CrossRef

Tinetti G et al (2016) The science of ARIEL. In: Proceedings of SPIE 9904, space telescopes and instrumentation 2016: optical, infrared, and millimeter wave, 99041X

Tritschler U, Cölfen H (2016) Self-assembled hierarchically structured organic–inorganic composite systems. Bioinspir Biomim 11:035002. https://doi.org/10.1088/1748-3190/11/3/035002
ADS CrossRef

Tsytovich V et al (2007) From plasma crystals and helical structures towards inorganic living matter. New J Phys 9:263. http://iopscience.iop.org/article/10.1088/1367-2630/9/8/263/pdf
CrossRef

Vincke K, Pfalzner S (2016) Cluster dynamics largely shapes protoplanetary disc sizes. MNRAS. arXiv:1606.07431

Visser P, van de Buit F (2015) Fourier spectra from exoplanets with polar caps and ocean glint. Astron Astrophys 579:A21
CrossRef

Vitasse O et al (2017) Interplanetary coronal mass ejection observed at STEREO-A, Mars, comet 67P/Churyumov-Gerasimenko, Saturn, and New Horizons en route to Pluto: comparison of its Forbush decreases at 1.4, 3.1, and 9.9 AU. J Geophys Res 122:7865

CrossRef

Wang H, Lineweaver C (2016) Proceedings of the 15th Australian space research conference. arxiv:1605.05003

Wolczek O (1982) New conceptions of unmanned planetary exploration – extra-solar planetary systems. Acta Astronaut 9:529
ADS CrossRef

Wong W et al (2016) Mimosa origami: a nanostructure-enabled directional self-organization regime of materials. Sci Adv 2(6):e1600417
ADS CrossRef

Yates J, Palmer P, Biller B, Cockell CH (2017) Atmospheric habitable zones in Y dwarf atmospheres. ApJ 836:184
ADS CrossRef

Zapatero-Osorio M-T et al (2016) Near-infrared photometry of WISE J085510.74-071442.5. Astron Astrophys 592:A80
CrossRef

Zarka PH, Treumann R, Ryabov B, Ryabov V (2001) Magnetically-driven planetary radio emissions and application to extrasolar planets. Astrophys Space Sci 277:293
ADS CrossRef

Zhu W et al (2016) Mass measurements of isolated objects from space-based microlensing. ApJ 825:60
ADS CrossRef

Zhu W et al (2017) An isolated microlens observed from K2, Spitzer and Earth. ApJ Lett. Submitted. arxiv:1709.09959

Zurbuchen TH (2017) NASA internal memo: next steps for WFIRST Program. Available at http://www.spaceref.com/news/viewsr.html?pid=50694